\documentclass[preprint,12pt,3p]{elsarticle}

\newcommand{\rqone}{\textbf{Which communication channels are adopted for safety analysis?}}
\newcommand{\rqtwo}{\textbf{How frequently used are safety analysis communication channels?}}
\newcommand{\rqthree}{\textbf{Which are the purposes of safety analysis communication channels?}}
\newcommand{\rqfour}{\textbf{What are the challenges when using safety analysis communication channels?}}

\usepackage[T1]{fontenc}
\usepackage[utf8]{inputenc}
\usepackage[english,american]{babel}

\usepackage{amssymb}

\usepackage{lineno}

\usepackage{booktabs} 

\usepackage{graphicx}
\usepackage{wrapfig}

\usepackage[flushleft]{threeparttable}
\usepackage{svg}
\usepackage{colortbl}
\definecolor{mygray}{gray}{.9}
\usepackage{multirow}
\usepackage{tcolorbox}

\journal{Journal of Systems and Software}

\begin{document}

\begin{frontmatter}

\title{Communication channels in safety analysis: An industrial exploratory case study}


\author[rvt]{Yang Wang\corref{cor1}}
    \ead{yang.wang@iste.uni-stuttgart.de}
    
\author[rvt]{Daniel Graziotin\corref{cor1}}
    \ead{daniel.graziotin@iste.uni-stuttgart.de}
    
\author[focal]{Stefan Kriso\corref{cor1}}
    \ead{stefan.kriso@de.bosch.com}
    
\author[rvt]{Stefan Wagner\corref{cor1}}
    \ead{stefan.wagner@iste.uni-stuttgart.de}
    
\cortext[cor1]{Corresponding author.} 
\address[rvt]{University of Stuttgart, Germany} 
\address[focal]{Robert Bosch GmbH, Germany}

\begin{abstract}
\emph{Context:} Safety analysis is a predominant activity in developing safety-critical systems. It is a highly cooperative task among multiple functional departments due to increasingly sophisticated safety-critical systems and close-knit development processes. Communication occurs pervasively. \newline
\emph{Motivation:} Effective communication channels among multiple functional departments influence safety analysis quality as well as a safe product delivery. However, the use of communication channels during safety analysis is sometimes arbitrary and poses challenges.   \newline
\emph{Objective:} In this article, we aim to investigate the existing communication channels, their usage frequencies, their purposes and challenges during safety analysis in industry. \newline
\emph{Method:} We conducted a multiple case study by surveying 39 experts and interviewing 21 experts in safety-critical companies including software developers, quality engineers and functional safety managers. Direct observations and documentation review were also conducted. \newline
\emph{Results:} Popular communication channels during safety analysis include formal meetings, project coordination tools, documentation and telephone. Email, personal discussion, training, internal communication software and boards are also in use. Training involving safety analysis happens 1-4 times per year, while other aforementioned communication channels happen ranging from 1-4 times per day to 1-4 times per month. We summarise 28 purposes of using these aforementioned communication channels. Communication happens mostly for the purpose of clarifying safety requirements, fixing temporary problems, conflicts and obstacles and sharing safety knowledge. The top 10 challenges are: (1) sensitiveness and confidentiality of safety analysis information; (2) fragmented safety analysis information; (3) inconsistent safety analysis information; (4) asynchronous channels; (5) a lack of tool support; (6) misunderstanding between developers and safety analysts; (7) language, geographic and culture limitations; (8) unwillingness to communicate (groupthink); (9) storage, authority, regulation and monitoring of safety analysis information; (10) a lack of documentation concerning safety analysis to support communication. \newline
\emph{Conclusion:} During safety analysis, to use communication channels effectively and avoid challenges, a clear purpose of communication during safety analysis should be established at the beginning. We have limitations primarily on the research context namely the scope of domains, participants and countries. To derive countermeasures of fixing the top 10 challenges are potential next steps.

\end{abstract}


\end{frontmatter}


\section{Introduction}
Safety analysis plays an important role in developing safety-critical systems. It aims at deriving safety requirements throughout the safety-critical system's development lifecycle \cite{martins2017requirements}. By performing safety analysis, potential causes of accidents can be identified, eliminated or controlled in design or operation before damage occurs \cite{leveson2011engineering}.
Due to an inherently complex nature of safety-critical systems, safety analysis concentrates not just on single component or function failures, but rather analysing from a system viewpoint including component interactions, cognitive complex human decision-making errors, social, organisational and management factors \cite{leveson2011engineering}. \par 

In safety-critical systems' organisation management, safety analysis is changing from individual tasks, which are mainly performed in the development team, to joint work with an alignment of multiple functional departments or external organisations, such as customers or suppliers \cite{martins2017requirements}. Safety requirements need to be provided by customers to the development team, while detailed safety information of products need to be provided from suppliers to original equipment manufacturers (OEMs). Among the stakeholders during safety analysis, communication occurs across different channels. \par

Leveson mentioned: \emph{``Risk perception (in safety analysis) is directly related to communication and feedback. The design of communication channels must be considered."} \cite{leveson2011engineering} (P.424). In the automotive domain, experts mentioned in the standard: \emph{``The organisation should institute, execute and maintain processes to ensure that identified functional safety anomalies are explicitly communicated to responsible persons"} \cite{iso26262} (Part 2, P.13). In the nuclear power domain, experts mentioned that: \emph{``Communication should be honest and open when answering to safety-related questions"} \cite{reiman2009evaluating} (P.53). In the aviation domain, experts mentioned: \emph{``Projects should plan for the appropriate level of technical oversight to ensure communication."} \cite{rierson2017developing} (P.528). It requires communication among team members, with the certification authority and customers. Therefore, in the safety-critical industries, the organisation management should promote effective communication channels during safety analysis at all levels. \par      

However, in practice, communication channels during safety analysis for safety-critical systems' development are uncultivated \cite{martins2017requirements}. Using formal documentation is a classic way during safety analysis to complement communication \cite{vilela2017integration}. It ensures a correct and complete information transfer. However, at present, the market of safety-critical systems is becoming fast-changing. The increasingly incoming customer requirements are captured by face-to-face communications \cite{hummel2013role}. These informal communication channels are arbitrarily used by practitioners and not regulated by relevant authorities. Thus, we support Keyton's \cite{doi:10.1146/annurev-orgpsych-032516-113341} and Kraut et al.'s \cite{kraut1990informal} views that an exploration of the existing communication channels during safety analysis in developing safety-critical systems becomes necessary, including their types, their usage frequencies as well as their purposes. \par

\begin{description}
\item[RQ 1] \rqone 
\newline Through investigating the existing communication channels during safety analysis, we can (1) have an overview of the communication scope in our research context during performing safety analysis; (2) identify the challenges on each communication channel, which will possibly influence the results of safety analysis, and further lead to an unsafe product. 
\item[RQ 2] \rqtwo
\newline Through investigating the using frequency of each communication channel, we can (1) acquire that which communication channel acts as a major role, which could have a large impact on the results of safety analysis; (2) help process improvements through adjusting the using frequency of  communication channels.  
\item[RQ 3] \rqthree 
\newline Through investigating the purposes of these communication channels, we can avoid the repeated use of multiple communication channels for the same purpose. In addition, for a clear communication purpose, we can select a suitable communication channel.
\end{description}

Furthermore, communication channels during safety analysis in developing safety-critical systems are problematic \cite{hummel2013role}. Practitioners in safety-critical industries notice their negative impacts from various perspectives, such as a poor management (an overwhelming amount of information and a loss of details) of safety analysis related information \cite{reiman2015principles} or little support for requirements engineering (a non-standardisation of nomenclature, the incompleteness of safety requirements and a lack of tool support) \cite{vilela2017integration}. 
Yet, as called by Dobson et al. \cite{lrsc}, Vilela et al. \cite{vilela2017integration} and Kraut et al. \cite{kraut1990informal}, scarce research explores the challenges in communication channels during safety analysis. \par    

\begin{description}
\item[RQ 4]  \rqfour
\newline Through investigating the challenges of these communication channels, we can provide practitioners hints of problems to avoid when they use one of the communication channels. Moreover, a clear research of challenges can help the researchers to look for suitable solutions.   
\end{description}

\subsection{Research Objectives}
In this article, the overall purpose is to investigate communication channels during safety analysis in developing safety-critical systems. First, we investigate the existing communication channels including their frequencies among multiple functional departments. Furthermore, we address their purposes in connection to the existing communication channels. Finally, we derive the top 10 challenges and map them with the aforementioned purposes.

\subsection{Contribution}
In this article, we have the following contributions: 
\begin{enumerate}
\item We find 9 popular communication channels including their usage frequencies during safety analysis. We notice that safety-critical systems have strict government regulations and the communication cannot be in public. Thus, the number of used communication channels is small in comparison with the existing communication channels in social media \cite{storey2010impact}. Yet, these 9 communication channels lay a foundation for further research. Most of them are used at a 1-4 times per week rate. 
\item We address 28 purposes based on the 9 communication channels during safety analysis. Although the number of communication channels during safety analysis is small, the purposes are diverse. 
\item We provide the top 10 challenges in the communication channels during safety analysis and map them with the 28 purposes. We highlight a possible way to select communication channels for achieving different purposes. The general challenges have specific manifestations as well as countermeasures when using different channels serving different purposes.    
\end{enumerate}

\subsection{Outline}
The article is organised as follows. In Section 2, we describe the background of communication channels and the related work regarding our study. We define a theoretical lens of communication channels during safety analysis in Section 3. Section 4 presents the case study design including context, research questions, data collection and analysis procedures. In Section 5, we report our study results. We discuss our implications and limitations in Section 6, before concluding our article in Section 7.

\section{Background and Related Work}
\subsection{Communication channels in traditional organisation management}
Communication channels in organisation management is a wide-ranging topic, which demonstrates how people use verbal and nonverbal messages to generate meanings within and across various contexts, cultures and media \cite{doi:10.1146/annurev-orgpsych-032516-113341}. \par
Keyton \cite{doi:10.1146/annurev-orgpsych-032516-113341} concluded a literature study by describing practical implications of areas concerning communication in organisation management. He summarised the theories from a philosophical perspective, such as a system theoretical approach, which depicts that messages are created, delivered and received by individuals in a complex web of relationships (safety analysis happens in a complex environment in industry), as well as a critical culture perspective, which points out that critical work encourages the exploration of alternative communication practices that allow greater democracy (safety analysis, as a critical factor, endures great influences from safety culture). In addition to the theories, the author posed future research questions, such as (1) \emph{What are effective ways for team members from different professions to develop shared meaning?} To perform safety analysis, a common understanding should be established among stakeholders. (2) \emph{How do culture of organisations and politics influence employees (mis)use of technology?} The execution of safety analysis is strongly influenced by safety culture and politics. In practice, communication channels are used with diverse taxonomies, such as verbal or nonverbal, synchronous or asynchronous and internal or external. \par
Johnson et al. \cite{johnson1994differences} investigated how to determine the use of communication channels in different taxonomies by conducting 380 surveys in a large midwestern state governmental agency. The authors mentioned that to use formal or informal communication channels, perceived applicability, output's effect and cultural norms are three possible criteria. Traditional safety-critical systems have a preference for formal communication to ensure preciseness \cite{bowen1993safety}. Yet, many informal communication channels arise currently in safety-critical industries. \par
Storey et al. \cite{storey2017social} focused on communication channels in organisation management during software development. They conducted a large-scale survey with 1,449 GitHub users to discuss the channels that the developers find essential to their work and gain an understanding of the challenges they face using them. 16 challenges are proposed under developer issues, collaboration and coordination hurdles, barriers to community building and participation, social and human communication, affordances, literacy and friction and content and knowledge management. Safety analysis intertwines with software development. Some of these challenges do exist during safety analysis in our context, such as geographic problems. Yet, they differ from manifestations. In software development, the authors considered geographic problems due to different time zones. An effective communication software, such as Slack, can reduce this problem. However, during safety analysis, geographic problems influence safety analysis information transmission and safety knowledge sharing. A pure communication software cannot solve this problem with respect to safety culture fundamentally. \par 
Kraut et al. \cite{kraut1990informal} investigated the use of informal communication channels in an R\&D organisation management through an experiment in which two small work groups were compared. The locations, the usage frequencies and the duration of informal communication are depicted with practical examples. They believed that in R\&D organisations, informal communication is keen to enhance collaborations and increase projects' success rates. During safety analysis, communication is becoming informal. Yet, these informal communication channels have not been considered as a way to improve collaborations.   \par

\subsection{Communication channels in modern organisation management}  
Modern software development processes are moving from heavy-weight with a huge amount of documents, to light-weight, with much oral communication. Communication channels receive much attention from practitioners \cite{cockburn2002agile} \cite{whitehead2007collaboration}. They advocate that face-to-face conversation is the best form of communication \cite{alliance2001agile}. The practical activities encompass pair programming \cite {williams2002pair} to enhance communication between developers or stand up meetings \cite{schwaber2002agile} to enhance communication among team members. Thus, the research concerning communication design or the importance of social skills are arising in modern organisation management. They are considered as a factor influencing team work quality and project success \cite{lindsjorn2016teamwork}. \par
Pikkarainen et al. \cite{pikkarainen2008impact} conducted a case study in F-Secure with two agile software development projects. The authors explored challenges focusing on internal and external communication. Since the groups are becoming self-organised in their research context, the communication among team members and between team members and customer, management, support group, enterprise staff show increasing challenges, such as a lack of coordination \cite{martini2016multiple} and a non-deterministic decision-making process \cite{conboy2017examining}. \par 
Hummel, Rosenkranz and Holten \cite{hummel2013role} conducted a systematic literature review with 333 relevant papers on agile software development and communication. They discussed the results from three directions: Input including team distribution, team size, project domain; Output including software development success; Agile software development including extreme programming (XP) and Scrum. The authors noted that the project domain poses specific issues on communication channels, such as safety or security-critical systems which rely on documentation, yet the changing requirements are not well documented. Moreover, when developing complex systems, there are lapses of memory which negatively influence communication. Safety-critical systems development is becoming light-weight \cite{staalhane2012application}. In particular, an integrated safety analysis happens more frequently. Special attention is needed in organisation. \par

\subsection{Communication challenges in safety-critical systems}
Developing or operating safety-critical systems encompasses specific communication challenges. Dobson, Moors and Norris \cite{lrsc} conducted a literature review concerning the communication in safety-critical systems by illustrating examples like tragic loss in an Esso gas plant explosion in 2001 (a combination of ineffective communication and inadequate hazard assessment), fire in a ship called Scandinavian Star in 1990 (a lack of communication between rescue co-ordination and passengers) and the Tenerife Airport Disaster in 1979 (misinterpretation by the Captain of the aircraft). The authors pointed out that (1) there is a link between communication and safety, (2) there is a range of mechanisms by which communication can fail, (3) there is a range of factors that shape the safety of communications, (4) formal, structured communication is most effective but needs to be use appropriately. 
Some possible communication problems contain missing, unnecessary, inaccurate, poor quality and ambiguous information. Improvements might be a careful specification, a moderate cutting-down, the utilisation of aids such as logs, the development of communication skills and the setting of standards for effective and safe communication \cite{prineas2010safety}. Some causalities have been investigated. \par
Flin et al. \cite{flin2008safety} classified these causalities into internal and external factors in their book. Internal factors can be attributed to characteristics of individuals, while external factors can be attributed to environments. Gibson \cite{gibson2007} identified five key elements by conducting a literature review, which include (1) the communication process of sender-medium-receiver, (2) the goals of the communication process, (3) the language used in the communication process, (4) the context of communication, and (5) individual factors. There are further recommendations in terms of explicitness, timing, assertiveness and active listening to improve communication in organisations \cite{flin2008safety}. These articles demonstrated a link between communication and safety-critical issues, as well as the specific elements and recommendations in safety-critical industries compared to the general communication in organisation management.    \par
Vilela et al. \cite{vilela2017integration} conducted a systematic literature review with 57 papers to investigate the communication between requirements engineering and safety engineering (including safety analysis). The research questions concern the activities, techniques, information artifacts, tools and benefits of performing safety analysis in connection with requirements engineering. 36 activities were listed by the authors. They mentioned that during these activities, a lack of a unified vocabulary among stakeholders hampers the communication. The techniques are illustrated by taxonomies, such as inductive/deductive and qualitative/quantitative. These taxonomies are considered to reduce the gap in communication between requirements engineering and safety engineering. The content of safety requirements is the main information through communication, which include safety-significant requirements and pure safety requirements as well as accidents, hazards, risks and harms. To promote an effective communication, practitioners should create an agreed-upon vocabulary and semantic structure containing all the relevant concepts, their relations and axioms within the safety domain for the purpose of exchanging information and facilitating reasoning. 66.67\% of the studies do not cite the use of tools. A lack of tools would lead to missing a sufficient guidance for execution and communication in safety analysis. Thus, it is worth noticing that an effective communication in safety analysis can reduce errors in requirements specification, improve system safety, help design, reduce costs and time, improve cooperation, enhance traceability, better present safety information, reduce workload, reduce interface faults, increase confidence and allow user feedback. 

\section{Communication in Safety Analysis}  

We define a theoretical lens of communication in safety analysis. We refer to the standards ISO 26262, ISO 14971 and IEC 60601 concerning the execution of safety analysis in safety-critical systems, as well as several internal safety analysis standards.  \par
Safety analysis happens primarily in four stages in the development of safety-critical systems. (1) At the beginning, a safety expert (he or she could be a functional safety manager, external safety expert or internal safety expert) together with other project members, who are responsible for facing customer projects, derive system-level safety requirements. Popular techniques are Hazard and Operability Analysis (HAZOP), Hazard Analysis and Risk Assessment (HARA) and Fault Tree Analysis (FTA). (2) After architectural design, the safety expert, development teams and other stakeholders derive detailed safety requirements and implement them in development. Popular techniques are Failure Mode and Effect Analysis (FMEA) for software and Failure Modes, Effects and Diagnostic Analysis (FMEDA) for hardware. (3) A verification, such as model checking, is necessary for testing the detailed safety requirements. (4) A validation, such as a review, is done by the deployment department or customers to test the system-level safety requirements. In addition to these four stages, safety analysis might happen during change impact analysis when there is a changing request. The concrete execution of safety analysis depends on whether they aim to develop a new product or reuse a product.   

\begin{figure*}[!h] 
\centering
\includegraphics[width=1.0\textwidth]{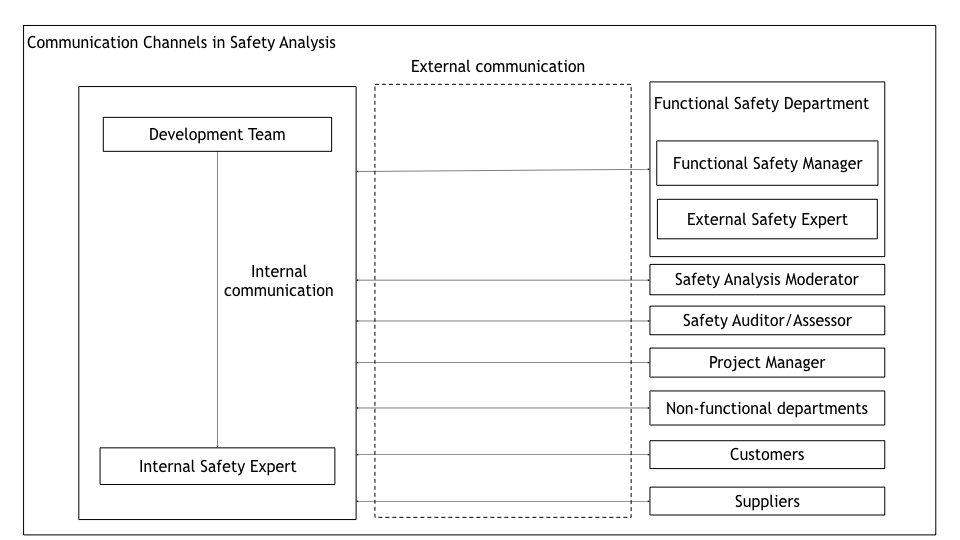} 
\caption{Theoretical lens of communication channels} 
\end{figure*}

The theoretical lens is shown in Figure 1. The communication encompasses internal communication and external communication. We simplify the roles in the \textbf{development team} with one \textbf{internal safety expert}, who mainly performs safety analysis, and other team members, which include architects, developers, testers and so on. The internal communication happens between the internal safety expert and other team members. Besides the development team, the industries also establish a \textbf{functional safety department}, which takes responsibility for the whole functional safety issues at the company level. In the functional safety department, a \textbf{functional safety manager} fixes mainly the external affairs and monitoring the execution of standards, while an \textbf{external safety expert} keeps contact with the internal safety expert to conduct training or knowledge sharing. In industry, the safety analysis related collaborative activities, such as execution and review, are guided by a \textbf{safety analysis moderator}, who is from a technical support department. The moderator ensures an official procedure to perform safety analysis. To ensure the safety assurance capability concerning process execution, the industries perform safety audit/assessment periodically by \textbf{a safety auditor or assessor}. It happens two to three times per year, especially when there is a deliverable product. The communication happens also between the development team and \textbf{project manager}, \textbf{customers}, \textbf{suppliers} and other \textbf{non-functional departments}, when the safety analysis issues concern project-level, product-level, purchasing or sales.

\section{Case study design}
We chose an exploratory case study design as proposed by Runeson et al. \cite{runeson2012case}. We conducted this case study in an inductive way by designing the theoretical lens in Figure 1. The reasons are: (1) There is no existing theory on communication channels during safety analysis, such as the possible channels and using regulations. (2) The exploration scope is indeterministic in terms of communication channels in safety analysis due to an omnicooperation. (3) The boundaries of safety analysis activities need to be clear for investigating communications, since some of them are blurred within development activities. This theoretical lens can provide us a clear boundary of our article and lay a foundation for our results.          
\subsection{Context}

\begin{table*}[!hbt]
\scriptsize
\centering
\caption{Research context}
\begin{tabular}{p{1.5cm}p{1.5cm}p{1.5cm}p{2cm}p{4.0cm}p{1.5cm}p{1.5cm}}
\toprule
Company  &  Size & Location & Domain & Main products & Employees & Participants   \\  \hline

A  & Large & Germany & Automotive & Automotive parts; Power tools; Electronics; Motorised bicycle motors. & 400,500 & 3  \\ 
A1 & Medium & China & Automotive & Automotive parts; Power tools; Electronics; Motorised bicycle motors. & 59,000 & 26  \\ 
A2  & Medium & China & Automotive & Gasoline engine management systems; Transmission control system; Hybrid and electric drive control system. & 9,400 & 19  \\ 
A3  &  Medium & China &  Automotive &Diesel systems for passenger cars, light and heavy commercial vehicles. & 1,800 & 3  \\ \hline
B  &  Large & Germany & Medical Equipment & CT/SPECT scanner; Angiography; X-ray products; Molecular diagnostics. & 372,000 & 4  \\  \hline
C  & Medium & Germany (Italy) & Automotive & Automotive lighting; Powertrain; Suspension systems; Motorsport. & 43,000 & 3    \\  \hline
D  & Small & Germany & Industrial 4.0 Based Product System & Technical strategy for production; ICT for manufacturing; Communication for factories; Process planning. & Less than 100 (unstable) & 2   \\
\toprule
\end{tabular}
\end{table*}

We conducted a multiple case study in seven safety-critical companies (three of which are subsidiaries of company A), as we can see in Table 1. We selected company A, together with A1, A2 and A3, as our biggest sample (with overall 51 participants). To cover different sizes of companies, we included company C (medium) and company D (small). To expand our research domains (company A is an international automotive industry), we included company B (an international medical equipment industry) and company D (a local industry 4.0 based production system). The duration of the investigated projects varies from 6 months to 3 years. The safety assurance processes of them follow ISO 26262 (automotive), VDA (automotive), Automotive SPICE, ISO 14971 (medical equipment), ISO 15004 (medical equipment) and IEC 60601 (medical equipment). The safety analysis is following the internal standards of FMEA, FTA and HAZOP. The investigated projects encompass functional development of ECU (Electronic Control Unit), NCU (New energy Control Unit) and body electronic control units, remote monitoring of CT machines and software services for smart factories. \par 
To include as many participants as possible, we encouraged the people in our research context, who have joined the safety analysis activities to participate in our survey. We selected 21 participants to join our interviews depending on their experience, positions and working years regarding safety analysis to gain the highest possible variability in our data. Other 39 participants, who are relative new to safety analysis, took part in the surveys. We asked the 21 interviewees before the interviews to ensure that they have not answered the surveys before. Their working experiences in safety-critical industries range from 1 to 23 years. As shown in Figure 2, most participants are quality assurance engineers and managers. 17 participants are from the quality assurance (QA) department. 10 participants are from the functional safety department. There are 23 managers and 3 experts. Other roles are 9 developers, 1 analyst and 2 leaders, as well as 1 participant in sales, 1 participant in purchasing and 1 participant in production department.

\begin{figure*}[!h] 
\centering
\includegraphics[width=1.0\textwidth]{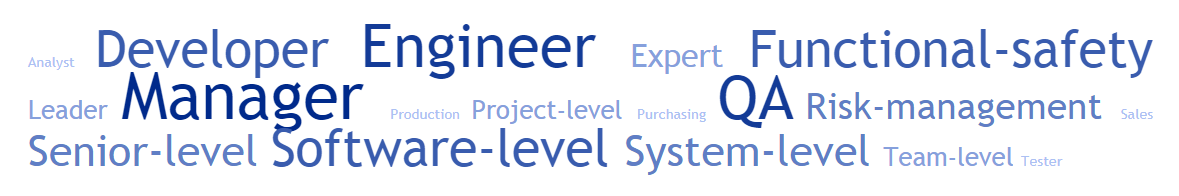} 
\caption{Participants} 
\end{figure*}

\subsection{Data collection}

As we can see in Figure 3, we conducted three rounds of data collection incrementally. In the first round, we ran surveys in seven companies between 2017-09-01 and 2017-10-31 to investigate the existing communication channels as well as their usage frequencies. In the second round, we conducted semi-structured interviews and documentation reviews in seven companies separately between 2017-09-01 and 2017-10-31 and between 2017-12-01 and 2018-01-31 to investigate their purposes and challenges. In the third round, from 2017-12-18 to 2018-01-05, the first author participated in several safety analysis meetings and the daily work in a functional safety department. The duration of participant observation is three weeks in company A2. 
Before data collection, we pre-interviewed several experts from four companies, either by telephone or by a face-to-face introductory meeting, to look through the organisation structure, decide on a common objective, establish agreements and help designing the surveys and interview questions. Each interview lasted one hour. These experts were further arranged to be the representative of each company. \par
In the first round, we used a survey to collect both qualitative and quantitative data, which cover the participant's background (positions, working years, the descriptions and durations of the running projects), the existing communication channels and the frequencies of using each communication channel. We sent the link to each representative via email, as well as the survey in electronic form to ensure that all the participants receive them. Since the investigation including the questions and predictable answers are sensitive in safety-critical systems, we decided to hand out the surveys through a company representative. This leads to a limited number of participants, which poses a challenge. During these two months, the first author checked the progress every two weeks through communicating with the representatives by videophone regarding the distribution of the surveys, as well as problems and feedback. \par
In the second round, we used semi-structured interviews to investigate the purposes and challenges of the communication channels. We selected firstly the communication channels from the results in the first round. We asked the interviewees to indicate possible additional communication channels as the first question. Moreover, we asked for the possible purposes of each communication channel. We went deeper to gain some real examples. The results may refer to product and customers' information, we keep them confidential. Lastly, we inquired about the challenges. We explained that the challenges they found could be specific to one communication channel, or in a general way across multiple channels. To make each challenge clear, we asked further in-depth sub-questions depending on participants' answers, such as \emph{``Who are the senders and receivers?"}, \emph{``What are the possible effects?"}, \emph{``How serious are the effects?"}, \emph{``What are the causal factors?"}, \emph{``How do you treat or fix this challenge?"}. We listed some possible in-depth sub-questions in the interview guide.       
Meanwhile, the first author reviewed the documents, which are related to the communication that happens during safety analysis, such as R\&D process instruments including safety analysis activities or interfaces in R\&D process, product development reports including the safety part of product development and quality management reports including safety quality management, quality management handbooks, R\&D risk management instructions, FMEA, FTA and HAZOP guidelines and working instructions, decision analysis and making reports and technical review reports. We inspected their contents, structures and relevant senders (editors) and receivers (readers or users) to analyse aspects such as \emph{``Are the contents fully written and could they be easy to understand by the receivers (readers and users)?"}, \emph{``Are they clearly structured by the senders (editors)?"} and \emph{``How long do the documents need to be looked through and be used?" }\par
In the third round, the first author conducted a direct observation in company A2. The first author took part in several team meetings such as a meeting concerning a comparison between the existing safety analysis procedure and a new version of ISO 26262, a meeting to discuss how safety analysis is executed in software and hardware development and how to coordinate with cyber-physical security. The first author observed the daily work in a functional safety department (she sat near to an internal safety expert who performed the safety analysis). Apart from the regular processes such as \emph{``How frequently does the internal safety expert communicate?"}, the first author observed also the internal safety expert's verbal communication with the functional safety manager, developers and stakeholders in non-functional departments, as well as the non-verbal communication, such as the internal safety expert's field notes, which recorded some outputs from communication. In the following section, we present how we analyse our data. 
 
\begin{figure*}[!h] 
\centering
\includegraphics[width=1.0\textwidth]{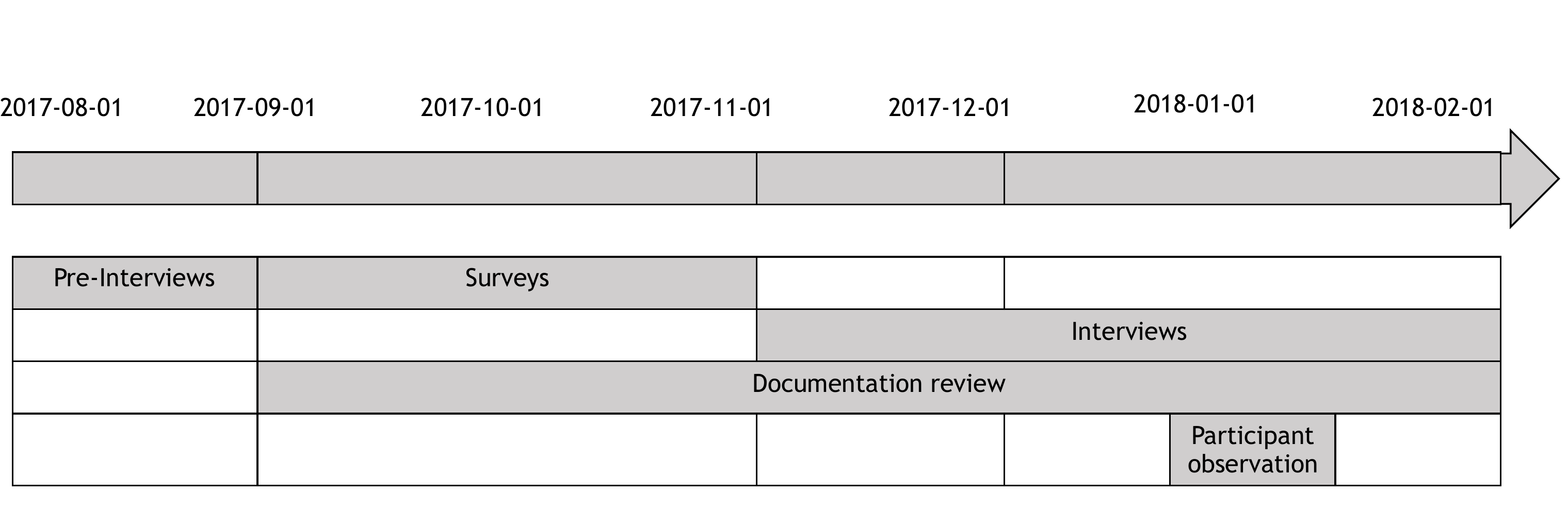} 
\caption{Timeline of data collection} 
\end{figure*}

\subsection{Data analysis}
To start answering our research questions, we investigate the participants' qualifications, such as positions, working experiences and running projects. \par 
For qualitative data, we consider basic coding steps and follow the logic of grounded theory coding \cite{strauss1994grounded}, since grounded theory is recommended as a powerful way for not only collecting and inspecting data, but also analysing data from the very beginning, as well as categorising and drawing meaningful conclusion. We chose the basic coding steps from Grounded Theory to systematically categorise and abstract the data. To start with, we conduct \textbf{open coding} to keep initial coding open-ended without having any preconceived concepts. Since the topic concerning communication channels has a wide range, even in safety analysis activities, open coding seems the most appropriate way to record transcripts line-by-line. Furthermore, based on an initial coding process, we get first categories. We list a preliminary category and conduct a \textbf{selective coding}. We focus on the codes relating to complementing the existing communication channels for RQ 1 and complementing the frequencies of using each communication channel, their purposes for RQ 3 as well as their challenges for RQ 4. The selective coding process is emergent and iterative. We compare the codes and refine our categories. For example, in terms of RQ 1, some interviewees mentioned that the communication channels lack a clear clarification, such as \emph{``If a personal meeting with two people is a formal meeting or personal discussion?"} We discuss the bias separately with an expert in company A to complement the results. In terms of RQ 2, after the first round of selective coding, some purposes show similarities, such as \emph{``We use email to inform a temporary meeting"} and \emph{``We call the colleagues to fix an emergent complaint"}, we group them as \emph{``A real-time notification"}. In addition, there is one sentence of code that encompasses two or more sub-purposes, such as \emph{``We go directly to the sales or deployment (the contact person) to organise a meeting when there comes a temporary but emergent customer's complaint"}. It is divided into \emph{``Fix customers' complaints"} and \emph{``Cooperate among multiple functional departments"}. We conduct a second round of selective coding to group similar purposes and divide mixed purposes. The same holds for RQ 4 concerning challenges. 
 
Lastly, to connect our results concerning the four RQs coherently, we conduct \textbf{axial coding} for RQ 1, RQ 3 and RQ 4. We link the existing communication channels with their purposes. We link the top 10 challenges with their purposes as well. We demonstrate an example of our coding phase including interview snippet, open coding, selective coding and axial coding in Table 2. 

\begin{table*}[!hbt]
\footnotesize
\centering
\caption{Example of coding phase}
\begin{tabular}{p{4cm}p{3.5cm}p{3.5cm}p{4cm}}
\toprule
\emph{Interview snippet}  & \emph{Open coding} & \emph{Selective coding \newline (possibly iterative)} & \emph{Axial coding}   \\  \hline

\textbf{Q}: ``What are the common purposes to facilitate communication concerning safety analysis?" \newline
\textbf{A}: ``...we have to exchange (safety analysis) information with our parent company...but the documents are always missing ... The functions are inherited, detailed architecture design documents are kept by them ... safety analysis cannot be done without considering interfaces with original products ..."  & We exchange safety analysis information with our parent company. The documents are missing. The functions are inherited. Detailed architecture design documents are kept by them. Safety analysis cannot be done without considering interfaces with the original products. & \textbf{Channels}: Documentation. \newline \textbf{Purposes}: Derive safety requirements. \newline Share knowledge. & CHANNEL\_Documentation \newline $<->$ \newline PURPOSES\_Derive safety requirements\_Share knowledge. \   \\  \hline

\textbf{Q}: ``Have you noticed some challenges in these communication channels?" \newline
\textbf{A}: ``...more importantly, we do care about the confidentiality. We categorised the data related to safety analysis with different confidential levels. High confidential data requires relevant authorities to read or transmit ..."  & We care about confidentiality problem. We categorise the safety analysis related information with different confidential levels. High confidential levels' data need authorities to read and transmit. & \textbf{Purposes}: Transfer safety requirements. \newline \textbf{Challenges}: Transmission of confidential information. \newline Authority problems.  & PURPOSES \_Transfer safety requirements \newline $<->$ \newline CHALLENGE\_Confidential information\_Authority problems. \   \\  \hline

\toprule
\end{tabular}
\end{table*}

For quantitative data concerning RQ 1, we choose pure numbers of participants to represent the utilisation of each communication channel. For quantitative data concerning RQ 2, in the first round, the participants scored the frequencies of the occurrences of each communication channel. The scale ranges from 1-4 times per day to 1-4 times per year. We consider that only the pure numbers in the surveys might show memory lapses \cite{hummel2013role}, the same as by asking the interviewees, a direct observation is necessary to validate the results. 
\section{Results}
\subsection{RQ1: \rqone}

\begin{figure*}[!h] 
\centering
\includegraphics[width=1.0\textwidth]{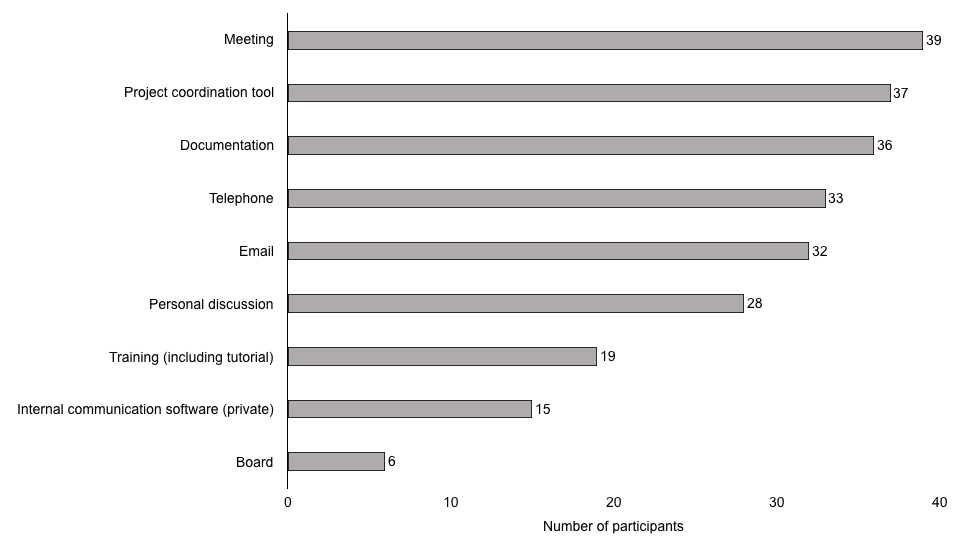} 
\caption{Communication channels in safety analysis} 
\end{figure*} 

As we can see in Figure 4, we find 9 communication channels during safety analysis, which are meetings, personal discussion, internal communication software (private), email, telephone, documentation, project coordination tools, training (including tutorial) and boards. Few participants mentioned the use of social communication software. However, considering that it is a rare and non-regulated case\footnote{We consider that it is due to a culture difference between Europe and Asia. More descriptions are shown in Section 6.2.}, we discussed with an expert in company A and decided not to include it in our results. We summarise the results as follows: \newline  
\textbf{Meeting:} All the 39 participants (100\%) mentioned meeting as a main communication channel during safety analysis. A meeting is defined as a gathering of two or more people that has been convened for the purpose of achieving a common goal through verbal communication, such as sharing information or reaching an agreement. It may take the form of face-to-face or virtually, as mediated by communications technology, such as a telephone conference call, a Skype conference call or a video-conference \cite{olson1992small} \cite{cutler2002distributed}. When performing safety analysis, the meetings include planning meetings, review meetings and requirements audit/assessment meetings. The joined members are invited by a meeting organiser, such as a safety analysis moderator. The communication might produce work products as outputs, such as relevant documentation including safety plan or safety requirements implementation decisions. \newline
\textbf{Project coordination tool:} 37 participants (94.9\%) mentioned using project coordination tools as a main communication channel during safety analysis. Project coordination tools aim to increase group awareness of current tasks and issues and provide a means for tracking progress and discussing next steps \cite{storey2017social}. Jira is the most popular one in our research context, together with the interfaces to other requirements management tools like Doors to keep the multiple levels and the traceability of safety requirements. Company A, company A1, company A2 and company A3 are using an Application Lifecycle Management (ALM) of IBM Rational Team Concert\footnote{https://arcadsoftware.de/produkte/rtc-rational-team-concert/}. In daily work, employees can trace the progresses of safety analysis transparently and provide feedback or comments timely. In other communication channels, such as meetings or personal discussion, project coordination tools provide an up-to-date information, such as the implementation of a safety requirement. \newline
\textbf{Documentation:} 36 participants (92.3\%) mentioned documentation as a main communication channel during safety analysis. According to the functional safety standards, safety analysis requires a large amount of documentation, such as safety analysis instrumentations, safety analysis execution plans and safety analysis audit/assessment reports. To record processes and results of safety analysis clearly, the employees have to rely on these documents. \newline
\textbf{Telephone: }33 participants (84.6\%) mentioned the use of the traditional telephone. Even though other modern communication software is springing up, most of the employees believe that telephone is more reliable for local communication and just as easy to use as in daily life \cite{baym2004social} \cite{kiesler1992group}. We obtained a novel channel calling \emph{a telephone call via Skype} in our research context. Company A links the telephone numbers with the Skype accounts. It enhances the work efficiency. For example, when someone is on the way among different work places, he or she can keep the communication via various mobile terminals, such as mobile phone or tablet personal computer.    \newline
\textbf{Email: }32 participants (82.1\%) mentioned the use of email. As an asynchronous communication channel, when the issues are not emergent and might involve not only one person, email provides time to structure the information for communication \cite{dabbish2005understanding}. Concerning safety analysis, email threads and the contents of emails are traceable. The traceability constitutes an advantage of using email. However, using email is not fully positive during safety analysis. Email can record the process of discussion, yet the practitioners during safety analysis do not prefer recording their discussion, rather only documenting their results. In addition, although using emails seems traceable, there needs a lot of efforts to search for the relevant information.   \newline
\textbf{Personal discussion: }28 participants (71.8\%) mentioned personal discussion as a communication channel during safety analysis. Personal discussion is a form of informal face-to-face communication. It happens spontaneously and less supported by communication technology \cite{kraut1990informal}. When performing safety analysis, an internal communication prefers mostly using personal discussions. It can ensure a correct understanding and a timely feedback. To perform safety analysis in modern software development, personal discussion needs attentions and the effectiveness of it will influence the quality of safety analysis to a great extent. \newline
\textbf{Training (including tutorial): }19 participants (48.7\%) mentioned training including tutorials as a communication channel. Training is a traditional way to enhance personal competence and share knowledge in industries \cite{saks2006investigation}. In terms of safety analysis, the education institutes sometimes lack such courses, especially the practical experiences in developing safety-critical systems. A training of safety analysis or functional safety standards is a normal way to cultivate employees. However, some of the employees are already experts from other relevant positions or departments in industries, such as product development or quality assurance departments. They have already a deep experience. Thus, not all the participants have to take part in the training. \newline 
\textbf{Internal communication software (private): }15 participants (38.5\%) mentioned internal communication software as a communication channel. We consider only the internal communication software that features a private chat, since not all the software featuring group or public chats in our research context (company A uses Skype that supports public chat, while company A2 uses Microsoft Office Communicator (OCS) that does not support public chat). It is a relative new communication channel popularising in the last decade. It integrates functions like a real-time notification, documents transformation and even meeting organisation \cite{storey2010impact}. Even though, the results indicate that internal communication software is not as much used during safety analysis as in other areas, such as software development \cite{storey2017social}. We speculate that concerning each single function, the employees  still have a better choice for achieving the purposes of safety analysis, such as for a real-time communication, telephone is more reliable to get in touch. \newline
\textbf{Boards: }6 participants (15.4\%) mentioned the use of boards. Boards, as communication channels, have been popularly used in industries and are becoming a tool during project management, such as whiteboards \cite{cherubini2007let}, in modern development processes. They are placed near the work areas. Some of them demonstrate the state-of-the-art or contributions in terms of safety analysis, such as a process flowchart of safety analysis in product development or the developed interfaces in safety analysis and relevant tools. In particular, when external experts or customers come to visit the company, boards are an intuitive way to show competitiveness.

\subsection{RQ2: \rqtwo}

Meetings are mostly held ranging from 1-4 times per week (18 out of 39 respondents, 46.2\%) to 1-4 times per month (18 out of 39 respondents, 46.2\%). Personal discussion happens 1-4 times per week (18 out of 27 respondents, 66.7\% ). Internal communication software is used mostly 1-4 times per week (11 out of 15 respondents, 73.3\%). Email (18 out of 32 respondents, 56.3\%) and telephone (23 out of 33 respondents, 69.7\%) are also frequently used 1-4 times per week. Documentation is written, read or managed 1-4 times per week (27 out of 36 respondents, 75\%). Project coordination tools are in use 1-4 times per day (36 out of 37 respondents, 97.3\%). Training is established 1-4 times per year with 100\% respondents' rate. Boards are possibly to be noticed 1-4 times per week (4 out of 6 respondents, 66.7\%).   
In summary, most of the communication channels (7 out of 9) are in use for safety analysis at least every week. Project coordination tools are in use for safety analysis everyday, while training (including tutorial) concerning safety analysis is mostly held every year. In particular, the same number of participants (18 participants) mention that the meetings are held between 1-4 times per week and 1-4 times per month. We discuss this point with an expert in company A and consider the reason to be the distribution of the company. For local projects, it is possible to establish necessary meetings. However, the distributed projects occupy also a large percentage in our research context. It is almost impossible for the employees, who are in distributed locations, to find a common time slot, such as in Germany, China, and USA. In these cases, meetings cannot be so frequently held at a 1-4 times per week rate and other communication channels are in use instead.             

\begin{figure*}[!h]
\centering
\includegraphics[width=1.0\textwidth]{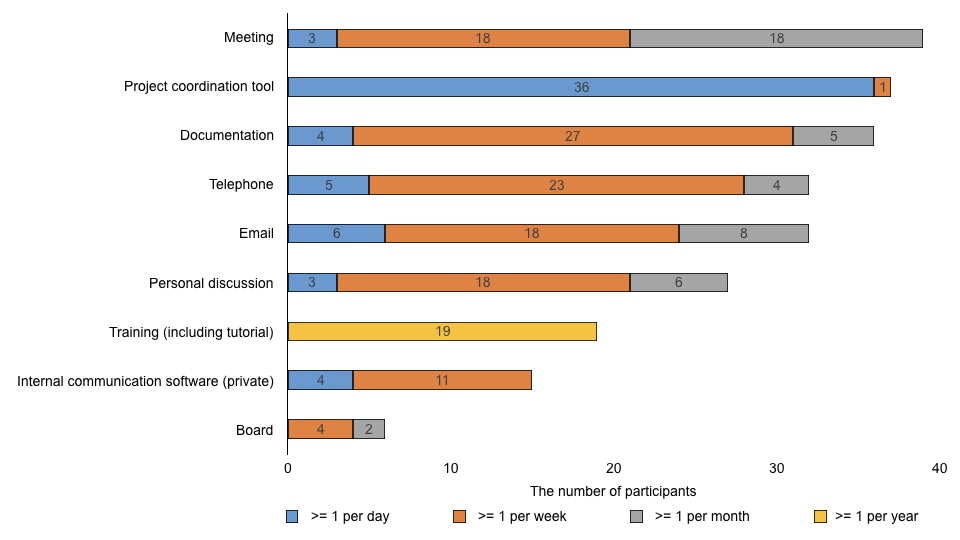} 
\caption{Usage frequencies of the 9 communication channel (The numbers in bars mean the number of participants)} 
\end{figure*}

\subsection{RQ 3: \rqthree}
We address overall 28 purposes of 9 communication channels during safety analysis, as we can see in Table 3. \newline
\emph{1. Transfer safety requirements (in).} To transfer safety requirements internally, the employees use meetings, documentation and project coordination tools. They discuss safety requirements in the form of meetings to keep it formal and possible to be established among multiple functional departments. They record the safety requirements in official documentation and project coordination tools to support daily communication. \newline 
\emph{2. Transfer safety requirements (ex).} To transfer safety requirements externally, meetings and documentation are implemented. However, contrary to the internal transfer of safety requirements, project coordination tools mostly have a limitation on permissions to external members. Some of the customers send the safety requirements per email, which seems to be non-regulated due to a lack of format of constructing safety requirements and an acceptance process of safety requirements. \newline 
\emph{3. Derive safety requirement.} A formal and thorough discussion is necessary to derive correct and complete safety requirements. Thus, meeting is the most effective way with respect to the number of participants and time. \newline 
\emph{4. Clarify safety requirements internally.} Many communication channels are applicable to clarify safety requirements internally, which include meetings, personal discussion, internal communication software (private), telephone, documentation and project coordination tools. It depends on the impact degree and scope of misunderstandings. A meeting is ideal for clarifying safety requirements with a severe impact. Personal discussion, internal communication software (private) and telephone are better for an individual clarification. Documentation is used to record an official clarification, while project coordination tools are specific for a clarification in an open mode. \newline 
\emph{5. Clarify safety requirement externally.} An official and reliable communication channel is extremely important to clarify safety requirements externally. Thus, meetings and documentation are the most appropriate channels. Email has also been used for an explanation at an early stage. \newline
\emph{6. Implement safety requirements.} The developers should implement safety requirements to their development. In terms of accountability, they prefer to execute it through meetings and documentation. The generation of ideas may happen through personal discussion. \newline
\emph{7. Trace safety requirements (bi).} According to the standards, safety requirements should satisfy a bi-directional traceability \cite{iec61508}. Project coordination tools give the employees a direct overview. However, to preserve them long-term, documentation is needed. In addition, some project coordination tools do not support multiple level safety requirements, such as Jira. \newline
 \emph{8. Planning.} To perform safety analysis planning, a planning meeting is a popular way, together with a safety plan as a work product. The execution process is shown in a project coordination tool. \newline
 \emph{9. Regular discussion.} It includes regular meetings in the development team or personal discussion among team members. Sometimes the team members use internal communication software. \newline
 \emph{10. Demonstrate periodic analysis results.} The safety analysis results should be demonstrated periodically to promote development. The results are transmitted automatically in the project coordination tools. The safety analysts demonstrate the results via the meeting and record them in the documentation. \newline
\emph{11. Demonstrate periodic V\&V results.} It is the same way with purposes 10. The acceptance criteria for safety verification and validation are recorded in the documentation and the review process is running in a meeting with a record in the project coordination tools. \newline
 \emph{12. Review.} There is a regular review meeting held at the end of a project. The results are recorded in the review report and in a project coordination tool at the same time. \newline
 \emph{13. Monitor the status.} Some practitioners facilitate communication for monitoring the status of safety analysis. The project coordination tool is the best way for monitoring the status in terms of a convenience and completeness overview. \newline
 \emph{14. Fix resources/supply problems.} Safety analysis sometimes faces the problems concerning a lack of resources or supply. The resources or suppliers are inside the company, yet among different subsidiaries or multiple functional departments. Meeting and personal discussion are the normal ways to fix such official but less frequent issues. \newline
\emph{15. Fix customers' complaints.} The employees receive the customers' complaints mostly through emails. Solving the problems is achieved through meetings. \newline
\emph{16. Establish commitments and make decisions.} The commitments are established among various roles, while the decisions are made in a formally strategic way. Meeting is the most suitable method. \newline 
\emph{17. Improve processes and techniques.} The processes and techniques of safety analysis are continuously developed. No matter how subtle the changes are, these changes have to be considered and discussed systematically and transparently \cite{pino2008software}. Meeting can gather omnifarious opinions and perform allocations from an overall perspective. \newline 
\emph{18. Fix temporary problems, conflicts and obstacles.} There are many unforeseeable and even tiny problems, conflicts and obstacles during safety analysis, possible communication channels are meeting, personal discussion, internal communication software, email and telephone. The selection of the channels depends on concrete issues. \newline
\emph{19. Cooperate among multiple functional departments.} A meeting is possible to be organised across multiple functional departments. The employees might also communicate through personal discussion, when they have a good relationship privately. The employees are also using internal communication software, email and telephone to communication with multiple functional departments. \newline
\emph{20. Help to understand the standards.} The activities and artifacts of performing safety analysis have to satisfy the standards. Not everyone has a solid background. To understand the standards, safety managers might introduce briefly in a meeting. More than that, the employees could ask relevant experts through personal discussion. To this end, training is popular in safety-critical companies to illustrate the standards systematically. \newline
\emph{21. Realise real-time notifications.} Some issues related to safety analysis are emergent, such as a serious customers' complaint, which might influence an ongoing delivery. In this case, the employees usually go directly to find the colleague (personal discussion), using internal communication software or telephone. Some project coordination tools provide a timely notification function in connection with email. \newline
\emph{22. Provide feedback and comments.} The employees are able to provide a timely feedback or comments through personal discussion, internal communication software, and telephone. Email is better for a clear explanation, while project coordination tools make the feedback and comments open to other employees. \newline
\emph{23. Enhance group cohesion.} Some communication channels among safety analysis and other stakeholders enhance safety-critical industries' cohesion, such as internal communication software and project coordination tools. Training is especially useful for new employees. \newline
\emph{24. Discuss off topics.} The contents of off topics might influence the safety analysis positively or negatively. The employees mainly use internal communication software or directly personal discussion. \newline
\emph{25. Share knowledge.} Knowledge sharing is considered important in safety-critical systems \cite{nesheim2014knowledge}, such as general safety knowledge or in-depth safety analysis information. It occurs through meeting, personal discussion, telephone, documentation, training and boards. \newline
\emph{26. Transfer documents or links.} The safety analysis related documents are mostly saved on a server. Yet, a small change of information management system as well as the complicated hierarchy of folders make some documents more difficult to be found or take more time than asking the responsible colleague. Thus, the employees might use communication channels to transfer them or their link. Internal communication software is a convenient way to share a link or an address, while email is able to send chunky files and keep track of the documents. \newline
\emph{27. Enhance safety culture.} Safety culture reflects the general attitude and approaches to safety and risk management \cite{leveson2011engineering}. Safety culture is difficult to evaluate \cite{guldenmund2000nature}. Documents are required by the standards. The satisfaction of such requirements of documents determines the organisation's safety assurance capability. It influences the safety culture. Responsibilities of specific roles, such as safety manager, are important to meeting the standards. Thus, a training of standards is necessary for remaining a good safety culture. Some participants mentioned boards as a good way to enhance safety culture through daily work, such as hanging out recent contributions. \newline 
\emph{28. Demonstration (external).} Boards are also popular to demonstrate the safety analysis capability of the organisations to external experts or customers. An intentional demonstration for a possible cooperation is conducted in meetings.   

\begin{table*}[!hbt]
\tiny
\caption{Purposes}
\begin{tabular}{p{6.2cm}p{0.8cm}p{0.8cm}p{0.7cm}p{0.5cm}p{0.6cm}p{0.6cm}p{0.8cm}p{0.7cm}p{0.7cm}}
\toprule
	& Meeting & Personal discussion & Csw.  & Email & Tel. & Doc. & Project coordination tool & Training & Boards \\ \midrule
	\rowcolor{mygray}
	1. Transfer safety requirements (in) & \checkmark &  & \ & \  & \  & \checkmark  & \checkmark  & \  & \  \\ 
	2. Transfer safety requirements (ex) &\checkmark &  &  & \checkmark  & \  & \checkmark  & \  & \  & \  \\ 
	\rowcolor{mygray}
	3. Derive safety requirements & \checkmark & \  & \  & \  & \  & \  & \  & \  & \  \\ 
	4. Clarify safety requirements (in) & \checkmark & \checkmark & \checkmark & \  & \checkmark  & \checkmark  & \checkmark  & \  & \  \\ 
	\rowcolor{mygray}
	5. Clarify safety requirements (ex) & \checkmark & \ & \  & \checkmark  & \  & \checkmark  & \  & \  & \  \\
	6. Implement safety requirements & \checkmark & \checkmark  & \  & \  & \  & \checkmark & \  & \  & \  \\ 
	\rowcolor{mygray}
	7. Trace safety requirements (bi) &  &  & \  & \  & \  & \checkmark  & \checkmark  & \  & \  \\ 
	8. Planning & \checkmark &  &  &  & \  & \checkmark  & \checkmark  & \  & \  \\
	\rowcolor{mygray}
	9. Regular discussion  & \checkmark & \checkmark & \checkmark &  & \  & \  &   & \  & \  \\
	10. Demonstrate periodic analysis results & \checkmark &  &  &  \  & \  & \checkmark  & \checkmark  & \  & \  \\ 
	\rowcolor{mygray}
	11. Demonstrate periodic V\&V results & \checkmark &  &  & \  & \  & \checkmark  & \checkmark  & \  & \  \\ 
	12. Review & \checkmark &  &  & \  & \  & \checkmark  & \checkmark & \  & \  \\ 
	\rowcolor{mygray}
	13. Monitor the status  &  & \  & \  & \  & \  & \  & \checkmark  & \  & \  \\ 
	14. Fix resources/supply problems & \checkmark & \checkmark & \  & \  & \  & \  & \  & \  & \  \\ 
	\rowcolor{mygray}
	15. Fix customers' complaints & \checkmark & \  & \  & \checkmark  & \  & \  & \  & \  & \  \\ 
	16. Establish commitments and make decisions & \checkmark & \  & \  & \  & \  & \  & \  & \  & \  \\ 
	\rowcolor{mygray}
	17. Improve processes or techniques & \checkmark &  & \  & \  & \  & \  & \  & \  & \  \\ 
	18. Fix temp. problems, conflicts and obstacles & \checkmark & \checkmark & \checkmark & \checkmark & \checkmark & \  & \  & \  & \  \\
	\rowcolor{mygray}
	19. Cooperate among multiple functional departments & \checkmark & \checkmark & \checkmark & \checkmark  & \checkmark  & \  & \  & \  & \  \\ 
	20. Help to understand the standards & \checkmark & \checkmark & \  & \  & \  & \  & \  & \checkmark  & \  \\ 
	\rowcolor{mygray}
	21. Realise real-time notifications & \  & \checkmark  & \checkmark  & \  & \checkmark  & \  & \checkmark  & \  & \  \\ 
	22. Provide feedback and comments &  & \checkmark & \checkmark & \checkmark & \checkmark  & \ & \checkmark  & \  & \  \\ 
	\rowcolor{mygray}
	23. Enhance group cohesion &  &  & \checkmark & \  & \  & \  & \checkmark  & \checkmark  & \  \\ 
	24. Discuss bordered or off topics &  & \checkmark & \checkmark & \  & \  & \  & \  & \  & \  \\ 
	\rowcolor{mygray}
	25. Share knowledge & \checkmark & \checkmark &  &  & \checkmark & \checkmark & \  & \checkmark & \checkmark  \\ 
	26. Transfer documents or links &  &   & \checkmark  & \checkmark  & \  & \  & \  & \ & \  \\ 
	\rowcolor{mygray}
	27. Enhance safety culture &  & \  & \  & \  & \  & \checkmark  & \  &\checkmark &\checkmark  \\
	28. Demonstration (external) & \checkmark & \  & \  & \  & \  & \  & \  & \  & \checkmark  \\ 
\toprule
\end{tabular}
\begin{tablenotes}
      \item Csw. is internal communication software. Tel. is telephone. Doc. is documentation. V\&V is verification and validation. In means internal, while ex means external. Bi means bi-direction.  
    \end{tablenotes}
\end{table*}

\subsection{RQ4: \rqfour}
\subsubsection{The Top 10 Challenges} 
We derive the top 10 challenges across 9 communication channels from interviews and check the results by a safety expert in company A. We list them as follows. \newline \newline
\fbox{\parbox{0.98\textwidth}{1. The communication of sensitive or confidential information should be monitored.}}\newline\newline
The results of safety analysis are mostly sensitive and confidential, since they encompass products in details, which influence a safe operation and end-use of products. Communication channels have to keep the information confidential, which includes ensuring that sensitive and confidential information is properly stored, maintained, secured, and accessible to those who need it \cite{banerjee2012ensuring} \cite{lucey2004management}. One interviewee mentioned: \emph{``We categorised the data (information) related to safety analysis with different confidentiality levels. Higher confidential data (information) requires relevant authorities to read and transmit. Yet, we do not regulate the verbal communication, including using social communication software to discuss the information. These information is not under monitor and control."} 
We note one factor concerning initiative leakage (a lack of regulations) of sensitive or confidential information through communication channels rather than passive leakage (a poor security and privacy assurance mechanism). As one participant mentioned: \emph{``We always avoid some non-regulated channels to transmit confidential and sensitive data (information), such as using e-mail to send safety requirements, even though some customers do this. We actually do not know clearly what is such regulations, just following our experiences."} To face this factor, organisation should establish relevant regulations on the communication channels, especially on the verbal channels, when performing safety analysis. The employees can understand clearly which sensitive and confidential information is able to transmit by which communication channels. Moreover, an alternative to verbal communication is recommended when performing safety analysis, such as text transmission.  \newline \newline 
\fbox{\parbox{0.98\textwidth}{2. Some safety analysis information is fragmented.}}\newline\newline 
The fragmentation will cause misunderstandings on safety analysis information. This happens particularly in text-based communication channels, such as safety analysis related-documentation and internal communication software. As one interviewee mentioned: \emph{``Mostly we can find out the safety-analysis relevant documents on the server. But sometimes the information is difficult to be understood or also possibly lacking some critical information, maybe not updated. The files are uploaded by authors themselves with no more reviews and even regular maintenance."} There is a specific role responsible for technical documents' maintenance in industries. However, the maintenance for online documents seems weak. Moreover, for the synchronous communication channels like internal communication software. One interviewee said: \emph{``To save time, there are too many personal abbreviations. That makes us confused and influences some emergent issues."}    \newline \newline
\fbox{\parbox{0.98\textwidth}{3. Some safety analysis information is inconsistent.}}\newline\newline
The tools such as project coordinate tools and safety analysis tools are updated frequently. However, other recordings do not keep the same space. As one interviewee mentioned: \emph{``In Jira and Doors, which we are using for recording safety requirements, there are all the updated safety requirements. But sometimes we use also doc files to record the results from safety analysis firstly and do not use those directly. The update (of official requirements documents) has a determined time point, which is not in a timely manner."} A real-time consistency seems important for the set of safety analysis related tools. In this paragraph, we focus on the contents, while the forth challenge focuses on the trigger time. \newline\newline   
\fbox{\parbox{0.98\textwidth}{4. Some communication channels concerning safety analysis are asynchronous, when they should be synchronous.}}\newline\newline 
Some communication channels are considered to be synchronous, such as Skype or telephone. However, they are not in our case. One interviewee mentioned: \emph{``When there is a temporary meeting, I will check if this colleague is online (Skype), then call him (or her). It works mostly. But once, I got an emergent customer complaint regarding our safety analysis, we need to find a colleague immediately, he was not online and could not be reached by telephone. I decided to go directly to his office and found him."} The asynchronous channels will hinder some emergent issues such as production and delivery and might cause fatalities. \newline\newline     
\fbox{\parbox{0.98\textwidth}{5. The communication channels concerning safety analysis lack tool support.}}\newline\newline 
Tools often put a limit to the strategies that can be adopted \cite{leveson2011engineering}. For effective communications, tools are important, especially the interfaces among different tools. When performing safety analysis, non-verbal communication takes an increasingly major part of it. An effective tool chain ensures the information's correctness, clearness, completeness and synchronous. However, the integration between safety analysis and functional development still lacks tool supports. Various organisational structures have various communication channels and tools. The new techniques are changing too fast to arrange an immediately effective use of them. As one interviewee said: \emph{``We start to use ALM (from IBM) for project management, but the interfaces (with the existing management tools) are not finished. The traceability is too poor."} The company has to consider an immediately developed interface with other safety analysis tools, when they introduce a new project management tool. Otherwise, when buying a new tool, a feasible interface to the existing toolchain should be considered by the company. As one interviewee mentioned: \emph{``APIS IQ-FMEA is a powerful tool that we used for performing FMEA. We have used it for many years and believe in it. Currently, we are considering to use a safety information and activity management tool. A lot of companies have introduced their tools. But our first criteria is if it has a feasible interface with APIS IQ-FMEA."}       \newline\newline 
\fbox{\parbox{0.98\textwidth}{6. Developers might misunderstand the information from safety experts.}}\newline\newline 
When the products to be developed have novel functions, there will be misunderstandings between functional developers and safety experts. One interviewee said: \emph{``It is no problem when developing old functions like ECU (electronic control unit). But these years we are starting a new functional module NCU (new energy control unit). The traditional safety experts, who have functional development experiences only on developing ECU, lack the new knowledge to understand detailed development."} The communication channels which support knowledge sharing might be helpful in this case.  \newline\newline
\fbox{\parbox{0.98\textwidth}{7. There are language, geographic and culture barriers in the communication channels.}}\newline\newline 
In terms of language, most of the companies in our context use English as the official language. Nevertheless, some professional terms concerning safety analysis are not uniform, such as incorrect verbs like \emph{operate safety analysis} instead of \emph{perform safety analysis}. This happens in the text-based as well as verbal communication channels. In terms of geographic, distributed locations interfere experience sharing. As one interviewee mentioned: \emph{``We send our colleagues, who perform safety analysis to other countries once a year to enhance communication and collaboration, but it works better for new employees. They want to learn new techniques and former experiences. For safety experts, we prefer that the execution of safety analysis should be suitable for our own environment or context. Different countries have different requirements and problems."} The basic knowledge has been well transmitted. However, the practical experiences are less shared and worse discussed. The knowledge sharing of safety analysis should not stop in the technique level, rather with more real-life cases. In terms of culture, distributed companies keep detailed safety analysis results and avoid to communicate them. There is a lack of trust among different cultures \cite{holmstrom2006global}. The parent company keeps some details, such as architecture design, of development and even safety analysis. When the subsidiaries inherit some function modules, they lack such details to perform a thorough and systematic safety analysis. Regulations among different countries might be a reason. \newline\newline    
\fbox{\parbox{0.98\textwidth}{8. The members from functional departments are unwilling to share safety knowledge with non-functional departments.}}\newline\newline
Functional (including functional safety) departments believe in their knowledge on development and safety of products. As one expert from the functional department said: \emph{``We don't see the necessity to do this."} We call this stereotype in groupthink theory\footnote{\scriptsize{Groupthink is a psychological phenomenon, which was introduced by Janis in 1971 \cite{janis1971groupthink}. It is a linear model of how seven antecedents (cohesion, group insulation, impartial leader, lack of norms, homogeneous, high stress from external threats, temporarily low self-esteem) increase the likelihood of premature concurrence seeking, which leads to eight psychological symptoms (illusion of invulnerability, collective rationalisations, belief in inherent morality of the group, stereotypes of out-groups, direct pressure on dissenters, self-censorship, illusion of unanimity, self-appointed mindguards).}}. This challenge exists during safety analysis and is specifically important for developing and ensuring a safe product \cite{wangw2018}. An interviewee said: \emph{``We do not think that other departments like purchasing or sales could help us a lot on developing a safe product. We know more and detailed on the product. For one project, we were required to train them with some general safety knowledge. We do not know how much it works. They also look unwilling."} In terms of safety knowledge sharing, the functional departments (including functional safety) should appreciate the needs and expectation of other job roles.  In addition, a lack of continuous learning and personal development processes reduce the passions of non-functional departments' employees to learn safety-analysis related knowledge, as one interviewee mentioned: \emph{``Safety analysis is not my main job, if just for this project and it does not help for future, we actually do not want to use too much time on it."} To this end, we should notice if the information is spread widely enough. As one expert said: \emph{``Sometimes the employees are forced to share the safety knowledge."} A lack of initiative makes the knowledge sharing becoming superficial. \newline\newline 
\fbox{\parbox{0.98\textwidth}{9. The storage, authority, regulation and monitoring problems on the transmitted safety analysis information.}}\newline\newline 
In terms of the storage of safety analysis information, one interviewee mentioned: \emph{``We have a central information management system on the server, but the safety analysis data (information) or files are not clearly classified. Some (of them) are mixed with other process documents."} The quality of transmitted information interleaves with the effectiveness of communication channels. The company should consider to establish a safety information management system to be separated with other information. In addition, the tools of information storage are important. An expert said: \emph{``The process of our data (information) storage is following the tools, not using tools to support process."} Thus, a suitable tool for safety information storage is needed, especially for modern safety-critical systems with a huge amount of data. As one interviewee mentioned: \emph{``The simulation data for autonomous driving system are thousands of giga bytes. It is relative difficult to find them when there is a migration happened 10 years before."} In terms of authority, one interviewee said: \emph{``In the information management system, we classify the files into different levels of authorities. But it is difficult to divide levels of safety requirements in Jira."} The authority should be noticed not only in the information management system, but also in the project management tool. In terms of regulation, some feedback are: \emph{``We are not clear about the regulations, just big sizes of files cannot be sent privately."} We notice that there are some hidden regulations on transmitting safety sensitive information. However, these regulations are not clearly and obviously announced. In terms of monitoring problems, the safety analysis information are monitored by an IT department as the same with other company level sensitive information. As one interviewee mentioned: \emph{``We do have a monitor group that monitors and controls the transformation of marked files. The high severity files have different kinds of mark. It could be traced by IT department."} The monitor of safety analysis information does work inside the company, but still needs enhancement. \newline\newline
\fbox{\parbox{0.98\textwidth}{10. There is a lack of technical documentation to support communication.}}\newline\newline
Safety analysis should be performed at the system level \cite{leveson2011engineering}. The detailed system description documents are necessary but sometimes not available. One interviewee mentioned: \emph{``Some products are developed on the basis of the original product. The functions are inherited. The detailed architecture design documents are kept by the original company or department. The execution of safety analysis cannot only be performed on new function modules without considering interfaces with original products. A systematic impact causes risks."} This might be caused by the culture (as we indicate in challenge 7), the authority problem (as we mentioned in challenge 9) of subsidiaries, or an incomplete and fragment record of safety analysis information (as we mention in challenge 2). Nevertheless, safety analysis related documentation should consider communication \cite{wang2017study}.   

\subsubsection{A Mapping between Purposes and Challenges}
\begin{table*}[!hbt]
\tiny
\caption{Purposes versus Challenges}
\begin{tabular}{p{6.2cm} p{0.60cm} p{0.60cm} p{0.60cm} p{0.60cm} p{0.60cm} p{0.60cm} p{0.60cm} p{0.60cm} p{0.60cm} p{0.60cm}}
\toprule
Purposes vs. Challenges	& 1 &  2 & 3  & 4 & 5 & 6 & 7 & 8 & 9 & 10 \\ \midrule
\rowcolor{mygray}
	1. Transfer safety requirements (in) & \checkmark &  & \ & \  & \  & \ & \ &  & \checkmark  &  \\ 
	2. Transfer safety requirements (ex) &\checkmark &  &  &  & \  &   & \  & \  & \checkmark  & \\ 
	\rowcolor{mygray}
	3. Derive safety requirements &  & \  & \checkmark  & \  & \  & \checkmark  & \  & \  & \ & \checkmark \\ 
	4. Clarify safety requirements (in) &  &  & \checkmark & \  &  & \checkmark  &  & \  & &   \\ 
	\rowcolor{mygray}
	5. Clarify safety requirements (ex) &  & \ &\checkmark &   & \  &   & \checkmark &\checkmark & \  &  \\
	6. Implement safety requirements &  &   & \  & \  & \  & \checkmark & \  & \  & \  &  \checkmark \\ 
	\rowcolor{mygray}
	7. Trace safety requirements (bi) &  &  &   &  & \checkmark  &  &   & \  & \  &  \\ 
	8. Planning &  &  &  &  & \  &   &  & \checkmark & \ &  \\
	\rowcolor{mygray}
	9. Regular discussion  &  &  &  &  & \  & \checkmark &   & \  & \  &  \\
	10. Demonstrate periodic analysis results & \ & \checkmark &  \checkmark&  \  & \  &   &   & \  & \ &   \\ 
	\rowcolor{mygray}
	11. Demonstrate periodic V\&V results &  & \checkmark  & \checkmark & \  & \  &   &   & \  & \  &  \\ 
	12. Review &  &  &  & \  & \  & \checkmark  &  & \  &  &   \\ 
	\rowcolor{mygray}
	13. Monitor the status  &  & \  & \  & \checkmark & \checkmark  & \  & \  & \  & \  &  \\ 
	14. Fix resources / supply problems &  & \checkmark & \  & \  & \  & \  & \  & \  & \  &  \\ 
	\rowcolor{mygray}
	15. Fix customers' complaints &  & \  & \  & \checkmark  & \  & \  & \checkmark  & \  &  &  \\ 
	16. Establish commitments and make decisions &  & \  & \  & \  & \  & \  & \  & \  & \checkmark & \checkmark \\ 
	\rowcolor{mygray}
	17. Improve processes or techniques & &  & \  & \  & \  & \  & \  & \checkmark & \  &  \\ 
	18. Fix temp. problems, conflicts and obstacles &  & &  &  & \checkmark & \  & \  & \  & \ & \\
	\rowcolor{mygray}
	19. Cooperate among multiple functional departments & &  & \ & \  & \ & \  & \  & \checkmark & \ & \checkmark \\ 
	20. Help to understand the standards &  &  & \  & \  & \  & \checkmark & \  &  & \checkmark  & \\ 
	\rowcolor{mygray}
	21. Realise real-time notification & \  &  &   & \checkmark  &  & \  &   & \  & \  & \\ 
	22. Provide feedback and comments &  &  &  & & \checkmark  & \ &   & \  & \checkmark & \\ 
	\rowcolor{mygray}
	23. Enhance group cohesion &  &  &  & \  & \  & \  & \checkmark  &  & \  & \\ 
	24. Discuss boards line or off topics & \checkmark &  &  & \  & \  & \  & \  & \  & \  & \\ 
	\rowcolor{mygray}
	25. Share knowledge &  & &  &  &  & \checkmark & \checkmark  &  &   & \\ 
	26. Transfer documents or links & \checkmark &   &   &  & \  & \  & \  & \ & \checkmark & \\ 
	\rowcolor{mygray}
	27. Enhance safety culture &  & \  & \  & \  & \  &   & \  &\checkmark &  & \\
	28. Demonstration (external) & \checkmark & \  & \  & \  & \  & \  & \  & \  & &  \\ 
\toprule
\end{tabular}
\begin{tablenotes}
      \item In. csw (private) is internal communication software. Tel. is telephone. Doc. is documentation. V\&V is verification and validation. In means internal, while ex means external.  
    \end{tablenotes}
\end{table*}

Based on the results from the interviews, we map the purposes with challenges, as shown in Table 4. The first line depicts the numbers of the top 10 challenges, while the first column depicts the purposes. \newline
Considering challenge 1: \emph{The communication of sensitive or confidential information should be monitored.} It happens often when the practitioners transfer and demonstrate safety analysis related information (purpose 1, 2, 24, 26, 28). The practitioners, who use the communication channels for achieving these purposes, should notice challenge 1. Other communication channels are better to avoid transferring sensitive or confidential information. \newline
Considering challenge 2: \emph{Some safety analysis information is fragmented.} The incomplete and fragment information occur through almost all the 9 communication channels, especially for demonstrating information (purpose 10, 11). Documentation, project coordination tools, email and internal communication software should keep safety analysis related information as complete as possible. When facing resource or supply problems, the provided information from them should notice completeness (purpose 14). \newline
Considering challenge 3: \emph{Some safety analysis information is inconsistent.} When the practitioners clarify the safety requirements (purpose 3, 4, 5) or demonstrate the results (purpose 10, 11), the communication channels should keep consistent information to avoid misunderstandings. Thus, the consistency among the recordings of meetings, the archived documentation and the project coordination tools seem important. \newline
Considering challenge 4: \emph{Some communication channels concerning safety analysis are asynchronous, when they should be synchronous.} It is extremely important when there is a need for real-time notification and timely monitoring (purpose 13, 15, 22). Thus, telephone, internal communication software should be able to reach a specific person when he/she is on-site. \newline
Considering challenge 5: \emph{The communication channels concerning safety analysis lack tool support.} A tool is necessary for tracing safety requirements (purpose 7), monitoring status (purpose 13) and providing feedback (purpose 22). The temporary problems necessitate tools to support recordings (purpose 18). We should arrange or design efficient tools and their interfaces as well. \newline
Considering challenge 6: \emph{Developers might misunderstand the information from safety experts.} The misunderstanding might occur between developers and safety experts when they use communication channels internally to derive, clarify, implement, discuss, review and share the safety analysis related information (purpose 3, 4, 6, 9, 12, 25). The understanding with respect to standards show also bias (purpose 20). The practitioners need to consider the understandability and an obstacle-free communication among multiple functional departments. \newline
Considering challenge 7: \emph{There are geographic, language and culture barriers in the communication channels.} When the industries aim to enhance group cohesion and share safety knowledge (purpose 24, 26), the communication channels, such as internal communication software, training and boards, might reduce the interferences from geographic, language and culture bias. To face customers or clarify safety analysis externally, it might happen in a multiple geographical distribution (purpose 5, 15). The relevant communication channels, such as meeting, should consider it. \newline
Considering challenge 8: \emph{The members from functional departments are unwilling to share safety knowledge with non-functional departments.} The communication channels regarding the cooperation among multiple functional departments (purpose 20) should avoid this challenge. For example, during meetings, a friendly environment for discussion is necessary. Personal discussion, such as by using internal communication software, email and telephone, is encouraged among multiple functional departments. When clarifying safety requirements externally (purpose 5), this challenge may happen due to a diverse knowledge background. When planning safety analysis (purpose 8), non-functional departments are keeping silence, as well as when improving processes or techniques (purpose 17). To enhance safety culture (purpose 27), non-functional departments should be included. \newline
Considering challenge 9: \emph{The storage, authority, regulation and monitoring problems on the transmitted safety analysis information.} To transfer safety requirements (purpose 1, 2, 26), storage, authority, regulation and monitoring are necessary. To establish commitment and make decisions (purpose 16), relevant regulations are needed. The fulfillment of standards should follow regulations (purpose 20). The feedback and comments need to be stored (purpose 22). \newline
Considering challenge 10: \emph{There is a lack of technical documentation to support communication.} The derive, implementation and decision making of safety requirements need a clear documentation to be understood by multiple functional departments and support an effective communication among them (purpose 3, 6, 16, 19).

\section{Discussion}

The main benefit of our article is that we investigate the general topic concerning communication channels in a concrete context concerning the execution of safety analysis. We aim to arouse the awareness of practitioners concerning safety analysis on their communication. The results are multiple: (1) we find 9 communication channels during safety analysis; (2) most of them happen 1-4 times per week; (3) we investigate 28 purposes of these 9 communication channels and (4) the top 10 challenges across these 9 communication channels.  \par

To compare with related work, Storey et al. \cite{storey2017social} found around twenty communication channels during software development. When performing safety analysis, the number of communication channels are smaller. In our article, we found nine communication channels during safety analysis to define a scope. We conjecture that (1) safety analysis is an activity within software development, (2) the confidentiality concerning sensitive safety-related data limit the number of possible communication channels, especially informal communication channels. \par
Kraut et al. \cite{kraut1990informal} mentioned that the usage frequency can show the importance of communication, which helps process improvements when performing safety analysis. In this article, we calculate the usage frequencies of the 9 communication channels during safety analysis to raise the importance of communication during safety analysis. \par
Vilela et al. \cite{vilela2017integration} proposed more than 11 pros that an effective communication in safety analysis can bring, such as reducing errors in requirements specification or helping design. However, to achieve an effective communication, the realisation of communication purposes is first and foremost. The existing research concerning communication purposes seem poor. We conjecture the reason might be that the general communication purposes distribute too wide-ranging. However, during safety analysis, the communication purposes can be accounted. \par
The challenges of communication concerning safety issues were mentioned in plenty of studies, which are either accident reports \cite{lrsc} or in terms of general organisation management \cite{hummel2013role}. The summarised communication challenges in software development are more than twenty \cite{storey2017social}. However, to the best of our knowledge, few research focus on the communication challenges, which might cause unsafe issues, during the execution of safety analysis in organisation management. In this article, we explore these challenges to provide hints of problems to avoid during safety analysis. 
          
\begin{tcolorbox}
\emph{``Being a good communicator is one thing. Knowing what to communicate is much more important." \cite{conboy2011people} (P. 52, Conboy et al.)}
\end{tcolorbox}

To see the results in this article, given the aforementioned opinion, we highlight the major contribution of our results as communication purposes \textbf{(RQ 3)}. During our research, a lot of the popular communication channels are used across various areas, not only for safety analysis. The challenges are summarised with similar manifestations, such as incomplete information and asynchronous implementation, but they are totally different in essence in terms of causalities, effects as well as solutions. Some of the challenges cannot map into safety analysis directly. We believe that a purpose during the communication in safety analysis determines why and what to communicate. Communication makes sense when the people achieve their communication purpose. \par

Thus, in this article, first, we map the 9 communication channels (RQ 1) with the 28 purposes. Within these 28 purposes, each one has 1 to 6 possible selections of communication channels to achieve it. For instance, to derive safety requirements (purpose 3), to establish commitments and make decisions (purpose 16) and to improve processes and techniques (purpose 17), the practitioners show only the use of meetings. To clarify safety requirements internally (purpose 4) and to share knowledge (purpose 25), each purpose has 6 communication channels as possible selections in practice. Second, we map the top 10 challenges (RQ 4) with their 28 purposes. The challenges rely heavily on their purposes and have different manifestations on each channel. For instance, ``the communication of sensitive and confidential information should be monitored" (challenge 1) when ``transfer safety requirements" (purpose 1, 2), as shown in Table 4, rather than whether the practitioners use meetings, email, documentation or project coordination tools, as shown in Table 3. \par

In addition to the major contribution on communication purposes, we demonstrate firstly the state-of-the-art in terms of the types \textbf{(RQ 1)} of the existing communication channels. The number is not huge comparing with existing communication channels in other areas such as social media \cite{storey2010impact}.  \par
\emph{Meeting} is a dominant one during safety analysis, which is able to achieve 20 out of 28 purposes. Especially, three purposes (3, 16, 17) can only be reached by meetings. The practitioners prefer to use meetings in traditional development processes which advocate formality as well as in modern development processes which aim to increase cooperation and communication. Meetings are possible for them both. \par
\emph{Project coordination tool}, \emph{documentation}, \emph{telephone} and \emph{email} are four popular communication channels in organisation management, as well as when performing safety analysis. They are able to reach 11 purposes, 12 purposes, 6 purposes and 7 purposes in safety analysis, respectively. \par
The importance of \emph{personal discussion} is an outstanding result in our study. 71.8\% participants mentioned that they are using it during safety analysis. It can achieve 11 out of 28 purposes. Based on our original conjecture, personal discussion might happen occasionally and especially rarely during safety analysis. Safety analysis is considered to be a technically in-depth activity, which needs an official communication or at least, more time to prepare. Comparing with other 8 communication channels, personal discussion seems to be the most unregulated one, which is extremely difficult to control, monitor and trace. We missed noticing its necessity. Yet, the results caught our attention.  \par
\emph{Internal communication software} is a novel channel since last decade. Even though it has been moderately mentioned in our context, the results show still its powerful functionalities that it can reach 9 purposes during safety analysis. \par
\emph{Training} is a concentrated way to enhance communication concerning safety analysis. It can reach 4 purposes. Apart from the basic training for new employees, some specific training programs, or even expert-level training, should be arranged. \par
\emph{Boards} are familiar for industries. It can achieve 3 purposes. Due to the modern development processes, such as whiteboards \cite{cherubini2007let} from Kanban, the practitioners could consider extending the boards' functionalities, particularly during safety analysis.  \par
To strength the existing communication channels further, the usage frequencies (\textbf{RQ 2}) in practice can demonstrate their importances. We set 4 scales including 1-4 times per day, 1-4 times, per week, 1-4 times per month and 1-4 times per year. Generally, 7 out of 9 communication channels are used 1-4 times per week. Project coordination tools are used 1-4 times per day. We discussed it with a safety expert and he said: \emph{``To start an (safety analysis) activity, we check the state as the first step, because we just switched from other tasks. Project coordination tool (Jira) is convenient, complete and intuitive."} A general overview is necessary before facilitating an issue. The project coordination tool is easy to use and it provides almost all the information or ways to get these information in a project. Training concerning safety analysis happens 1-4 times per year. From our viewpoint, it is reasonable.             
 
As an initial step to investigate communication channels during safety analysis, we concentrate on the top 10 challenges (\textbf{RQ 4}) based on analysing our qualitative data and a final check by an expert in company A. Several of them show similar problems as in general communication channels, such as incomplete and fragmented information. Yet, during safety analysis, we notice that the fragmentation of information are in high severities and may cause fatalities. The completeness is not dominant. As one expert mentioned: \emph{`` 90\% (incomplete) information sometimes is fully sufficient for solving the problems during safety analysis. But when the information is complete but fragmented, such as safety cases, which are spread over different tools, that seems serious."}  Geographical problems are also popular in general communication channels. Developers cannot get in touch to exchange states. However, during safety analysis, it influences not only the information transmission, but also safety knowledge sharing. These challenges need separate considerations. Other challenges are specifically existing in the communication channels during safety analysis, such as confidentiality concerning safety analysis information and groupthink including unwillingness to share safety analysis information with non-technical members.   \par

\subsection{Implications}
For \emph{theory}, RQ 1, RQ 2 and RQ 3 might provide implications. As far as we know, there are no reported results on the existing communication channels as well as their usage frequencies and purposes during safety analysis. An overview of communication channels' scope (Figure 1 and Figure 4) helps a further investigation on challenges. The using frequency of each communication channel (Figure 5) shows the importance of each one and helps further process improvement when performing safety analysis. A clear purpose (Table 4) can help practitioners selecting a suitable channel during safety analysis. We believe that in developing safety-critical systems, the practitioners are not keen on enriching the amount of channels, rather, solving the challenges on the existing communication channels to ensure a safe development process and a safe product. Thus, the 9 communication channels during safety analysis are convincing. Most of the communication channels are frequently used per week. In addition, RQ 3 provides 28 purposes of communication during safety analysis. In the future, we expect that researchers could expand this study to more safety-critical companies, domains and countries to check and enrich our initial results on the number of communication channels, their usage frequencies and purposes. Depending on the sizes of companies, there might be various answers. \par
For \emph{practitioners}, RQ 4 might arouse more interest. Practitioners in safety-critical industries may have these problems in their running projects. The results of RQ 4 (Table 5) can provide practitioners hints of problems to avoid when they use one of the communication channels. However, we do not provide any solutions to each specific challenge in this article, since we believe that the contexts and tools of each communication channel are changing, the challenges may remain challenges in the future but with different manifestations. The solutions should be derived in a specific way. Moreover, the mappings between RQ 1 and RQ 3 as well as between RQ 3 and RQ 4 seem practically useful. For achieving the communication purposes during safety analysis, the practitioners can select the possible channels from Table 3 and know the possibly relevant challenges from Table 4. In the future, we expect that the practitioners could propose more challenges that they have met as well as solutions, either in a general way or specifically in a context.   

\subsection{Limitations}

We believe that three major limitations threaten the results of our study.

Our sample might not cover all possible roles during safety analysis; the results could be biased by an over-representation of medium sized companies, and we could not cover all possible domains where safety analysis is critical (e.g., aerospace). Finally, the sample could cover only companies from three countries. We believe, however, that the sample is rich and meets a high standard for qualitative case studies. Our sample of 60 participants covers all possible company sizes (small, medium, and large), offers data from two geographically and culturally diverse zones (Germany/Italy and China), and considers three different domains of interest (automotive, medical, and industry 4.0). For a stronger generalization of our results, we call for survey studies between companies to \textit{go wide} where we could \textit{go deep} within companies.

Communication occurs frequently and often spontaneously, so it is challenging to observe it directly. 
In our case study, while we could perform direct observation sessions, we mostly collected perceptions and experiences of participants. Memory recalling and other cognitive biases could over or under-represent certain communication channels and their usage frequency. This is a typical issue of observation studies, questionnaires, and interview-based designs \cite{Creswell2009} that we wish to recall nonetheless.

Furthermore, causality chains (e.g., those in RQ 3 and RQ 4) are harder to empirically demonstrate as there was no controlled experiment testing the claims. There is an open debate on whether causality can be inferred from research approaches other than controlled experiments \cite{Creswell2009,Djamba:2002jo,Glaser:2013ha}. We agree with several authors, e.g., Gl{\"a}ser et al. \cite{Glaser:2013ha}, take the stance that qualitative data analysis can be used to infer causality from the experience of participants, provided that there is a strong methodology for data gathering and analysis. We claim that our methodology is robust. We have employed data triangulation validation whenever possible, for example by adding a three weeks long direct observation to validate the results and by offering our results to a senior functional safety expert from company A to check. Still, given the lack of prior literature, we deem our study to be exploratory in its nature, not confirmatory. We call for future research to empirically demonstrate the relationship chains that we uncovered.
      
\section{Conclusion and Future Work}
To investigate the communication channels during safety analysis, we conducted an industrial exploratory case study in 7 safety-critical industries with overall 60 participants. We used surveys, interviews, documentation review and participant observation in a large automotive company (with three medium subsidiaries), a large medical equipment company, a medium automotive company and a small industrial 4.0 based production line company. During three rounds of data collection, we found 9 communication channels during safety analysis. Most of them happen 1-4 times per week. We summarised 28 purposes of these 9 communication channels and the top 10 challenges across these 9 communication channels. We notice the importance of communication purposes and map them with the existing channels and challenges. Our study is mainly limited by the domains, participants' background and countries, which cover only 3 safety-critical domains and mainly in Europe and Asia with only 60 participants. \par
Further research could expand our study and enrich our preliminary results, such as with more subjects to generate communication channels between different senders and receivers, in more safety-critical domains like aviation or railway industries. In addition, a quantitative assessment for these challenges and some practical solutions are expected to be generated.    \par

\section{Acknowledgements}

We would like to thank all the participating companies as well as colleagues helping out in different parts of this research, especially Yun Jiang, Liqiang Dong, Jakubasch Ronny, Schmidt Klaus, Michael Fijala, Roberto Secchi and Suneet Jain, as well as their help to our previous research on groupthink in safety analysis \cite{wangw2018}.

\section{Funding}

Yang Wang has been supported by \emph{LGFG Fellowship (Promotionsstipendien nach dem Landesgraduiertenf{\"o}rderungsgesetz, Baden-W{\"u}rttemberg)}. Daniel Graziotin has been supported by the Alexander von
Humboldt (AvH) Foundation.  





\bibliographystyle{elsarticle-num}
\bibliography{sigchi}
\vspace{2.5cm}






 \begin{wrapfigure}{l}{50mm} 
    \includegraphics[width=1.4in]{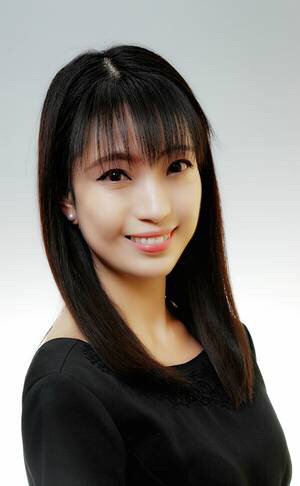}
  \end{wrapfigure}\par
  \textbf{Yang Wang} finished her PhD study in computer science at the University of Stuttgart from  April 2015 to September 2018. Her research interest is on agile methods in safety-critical systems. She starts her research from concentrating on the use of Systems-Theoretic Process Analysis (STPA) for safety and security analysis. In addition, her research interests include agile testing, agile project management and organisation culture in safety-critical systems. Yang was awarded an LGFG (Promotionsstipendien nach dem Landesgraduiertenf{\"o}rderungsgesetz, Baden-W{\"u}rttemberg) Fellowship. Before, she has practical working experiences at Robert Bosch GmbH and Siemens AG. Contact her at yang.wang@iste.uni-stuttgart.de.   \newline \newline

 \begin{wrapfigure}{l}{50mm} 
    \includegraphics[width=1.8in]{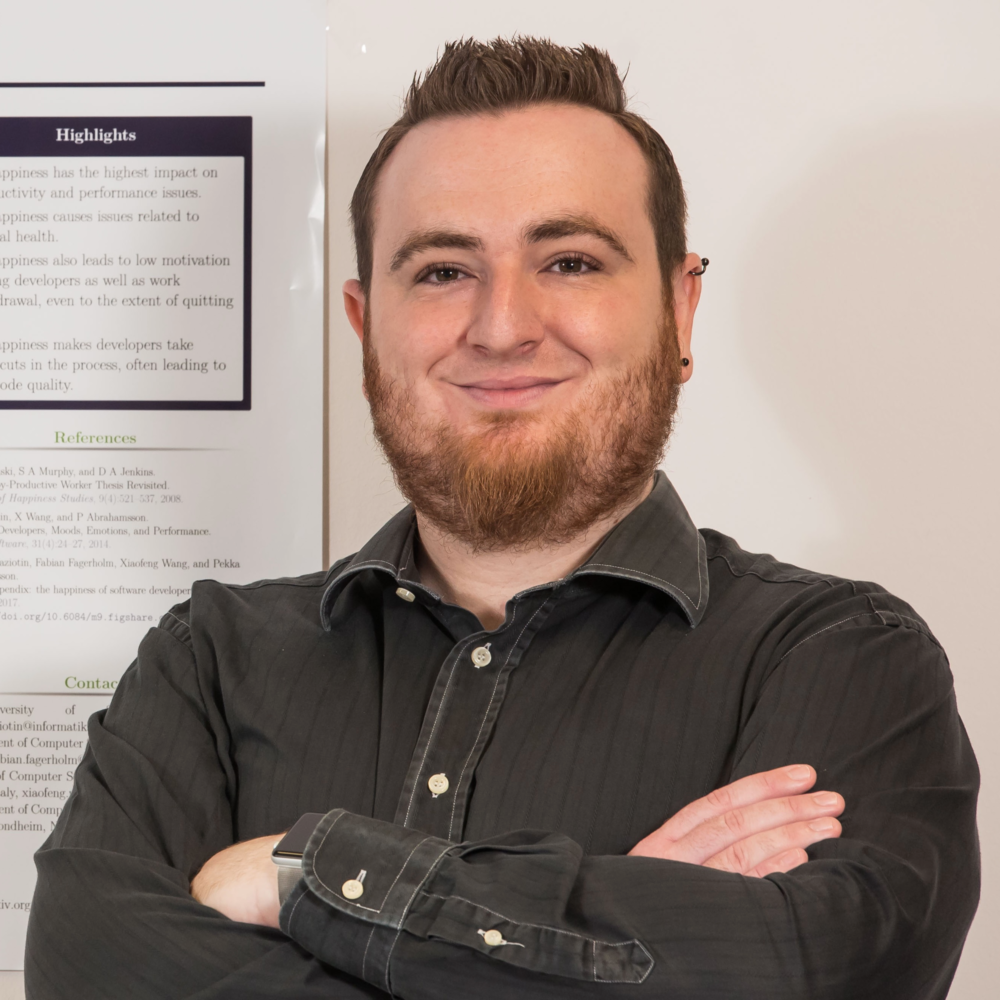}
  \end{wrapfigure}\par \par \par
   \textbf{Daniel Graziotin} is a postdoctoral researcher at the University of Stuttgart, Germany. His research interests include human, behavioral, and psychological aspects of empirical software engineering, studies of science, and open science. He is associate editor at the Journal of Open Research Software and academic editor at the Research Ideas and Outcomes (RIO) journal. Daniel was awarded an Alexander von Humboldt Fellowship for postdoctoral researchers in 2017, the European Design Award (bronze) in 2016, and the Data Journalism Award in 2015. He received his PhD in computer science at the Free University of Bozen-Bolzano, Italy. Find him on Twitter at @dgraziotin.  \newline \newline \newline \newline 
  
   \begin{wrapfigure}{l}{50mm} 
   \includegraphics[width=1.8in]{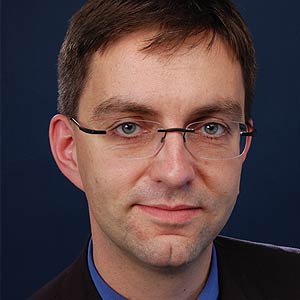}
  \end{wrapfigure}\par
  \textbf{Stefan Kriso} is the head of the Bosch \emph{Center of Competence Functional Safety}. Within the Corporate Research and Advanced Engineering area he led from 2004 till 2011, the task was the representation of Bosch in the national and international standardization boards of ISO 26262 and the coordination of the Bosch-wide rollout of ISO 26262. He is member of the German ISO 26262 working group (VDA NA052-00-32-08-01 AK) and leader of the ZVEI Ad-hoc working group ISO 26262. Contact him at stefan.kriso@de.bosch.com.   \newline \newline \newline

     \begin{wrapfigure}{l}{50mm} 
    \includegraphics[width=1.6in]{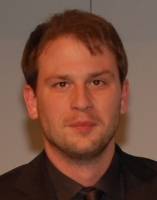}
  \end{wrapfigure}\par
  \textbf{Stefan Wagner} is a full professor of software engineering at the Institute of Software Technology, University of Stuttgart. He studied computer science in Augsburg and Edinburgh and holds a doctoral degree in computer science from the Technical University Munich. His research interests include requirements engineering, software quality, safety and security engineering, agile software development and empirical software engineering. He is a member of GI, ACM and IEEE. Contact him at stefan.wagner@iste.uni-stuttgart.de. \par

\end{document}